\documentclass{PoS}
\usepackage{epsfig,amsmath,amssymb,bm}

\newcommand{\ba}{\begin{array}}
\newcommand{\ea}{\end{array}}

\newcommand{\be}{\begin{equation}}
\newcommand{\ee}{\end{equation}}
\newcommand{\bea}{\begin{eqnarray}}
\newcommand{\eea}{\end{eqnarray}}

\newcommand{\sun}{{\odot}}
\title{Dense QCD and phenomenology of compact stars 
}

\ShortTitle{Massive compact stars}

\author{\speaker{Armen Sedrakian}\\
Institute for Theoretical Physics,
J. W. Goethe University,\\D-60438 Frankfurt am Main, Germany\\
        E-mail: \email{sedrakian@th.physik.uni-frankfurt.de}}

      \abstract{ I discuss three topics in physics of massive (two
        solar-mass and larger) neutron stars containing deconfined
        quark matter: (i) the equation of state of deconfined dense
        quark matter and its color superconducting phases, (ii)
        the thermal evolution of stars with quark cores, (iii)
        color-magnetic flux tubes in type-II superconducting quark matter
        and their dynamics driven by Aharonov-Bohm interactions with
        unpaired fermions.  }

\FullConference{Xth Quark Confinement and the Hadron Spectrum,\\
		October 8-12, 2012\\
		TUM Campus Garching, Munich, Germany}

\begin{document}

\section{Introduction}

Compact star phenomenology offers a unique tool to address the
outstanding challenges of the modern particle and nuclear physics, in
particular the phase structure of dense quantum chromodynamics (QCD).
The integral parameters of compact stars depend on their equation of
state (hereafter EoS) at high densities. Measurements of pulsar masses
in binaries provide the most valuable information on the underlying
EoS because these, being deduced from binary system parameters, are
model independent within a given theory of
gravity~\cite{Kramer:2012zz}.  The recent discovery of a compact star
with a mass of 1.97~$M_{\odot}$, which sets an observationally
``clean'' lower bound on the maximum mass of a compact star via the
measured Shapiro delay~\cite{Demorset:2010}, spurred an intensive
discussion of the phase structure of dense matter in compact stars
consistent with this limit~\cite{MassiveStars,Bonanno:2011ch}.  Pulsar
radii have been extracted, e.g., from modeling the X-ray binaries
under certain reasonable model assumptions, but the uncertainties are
large~\cite{Radii}.  A recent example is the pulse phase-resolved
X-ray spectroscopy of PSR J0437-4715, which sets a lower limit on the
radius of a 1.76\, $M_{\odot}$ solar mass compact star $R > 11$ km
within 3$\sigma$ error (Bogdanov, in Ref.~\cite{Radii}).

Large masses and radii are an evidence for the stiffness of the EoS of
dense matter at high densities. Commonly, phase transitions, e.g.  the
onset of quark matter, softens the EoS close to the transition point,
and therefore reduces the maximum achievable mass of a
configuration. Thus, if quark matter exists in compact stars, there
should be a delicate balance in Nature which allows large enough
densities in compact stars to achieve quark deconfinement and stiff
enough EoS of strongly interacting matter to support large masses and
radii~\cite{MassiveStars}.  Indeed, as I review below, one can
construct stellar configurations that are dense enough for 
baryons to deconfine into quarks and are massive enough to be
consistent with the maximum mass bound mentioned above. These
configurations have also radii large enough to be consistent with the
existing bounds ~\cite{Radii}.

The information gained from measurements of the gross parameters of
compact objects, such as the mass, radius  or the moment of inertia give us
only indirect information on potential phases of matter in their
interiors. The studies of rotational, thermal and magnetic evolutions
of neutron stars provide complementary information, which is sensitive
to the composition of matter.  Indeed, modeling cooling neutron stars
requires a detailed knowledge of quantities such as the specific heat
of components of matter and its neutrino emissivity, which are
strongly composition dependent.  Thus, the cooling of  massive neutron
stars needs to be consistent with the ideas of formation of quark
matter with stiff equations of state.  The studies of cooling of such
stars reveals the complexity of the problem~(for the recent work
see Refs.~\cite{Alford:2004zr}).  Furthermore, the recent
observation of the substantial change in the temperature of the
neutron star in Cas A poses a challenge for the theory to explain
drastic short-term drop in the temperature of this neutron star~(for a
review see Ref.~\cite{Shternin:2010qi}). Magnetic evolution of a
compact star will also depend on the structure of magnetic field in
the quark phases, which in turn depends on the dynamics of quantum
vorticity in the quark superconductor or
superfluid~\cite{QV,Alford:2010qf}. For example, if the 
flux-tube dynamics in the type-II superconducting phases
of the star occurs on evolutionary time-scales, it will directly affect the
magnetic field components of the star that lead to its secular spin-down.

\section{Two solar-mass compact stars with hyperons and color superconducting quarks}
\label{sec:sup_phases}

Hybrid stars, i.e. baryonic stars containing quark cores, with two
solar masses have attracted a lot of attention in recent years. Cold
quark matter should be in one of the possible superconducting phases
due to the attractive component in the gluon exchange quark-quark
interaction~\cite{CSCGeneral}. These two aspects - high masses and color
superconductivity - appear to be minimal ingredient of realistic
hybrid star modeling. Here I describe briefly a recent work on a 
model~\cite{Bonanno:2011ch}, which exemplifies the key features 
of this class of compact stars.

The nuclear EoS, as is well known, can be constructed starting from a
number of principles. In Ref.~\cite{Bonanno:2011ch}  relativistic
mean-field models were employed to model the low density nuclear
matter.  As is well-known, these models are fitted to the bulk
properties of nuclear matter and hypernuclear data to describe the
baryonic octet and its interactions~\cite{LBL-30645}.  Among
the existing parameterization the stiffest EoS is achieved with the
NL3 and GM3 parameterizations.  The high-density
quark matter was described in Ref.~\cite{Bonanno:2011ch} by an NJL
Lagrangian, which is extended to include the t' Hooft interaction term
 and the vector interactions with a coupling $G_V$.  
\begin{figure}[tb]
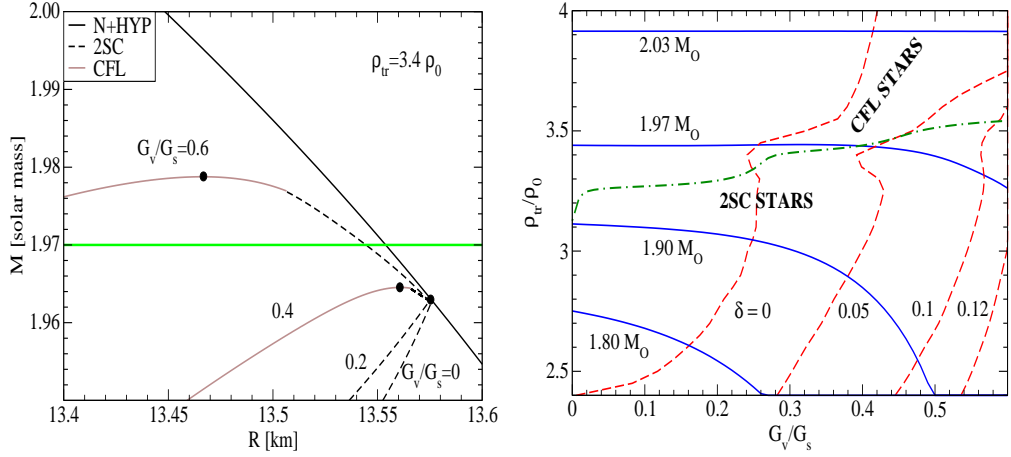

\hskip 0.1 cm
\begin{center}
\epsfig{figure=M_R,height=6.cm,width=6.5cm,angle=0}
\hskip 0.2cm
\epsfig{figure=parameter_space2,height=6.cm,width=6.5cm,angle=0}
\vskip 0.3cm
\caption{ {\it Left panel.}~Mass vs radius for configurations with
  quark-hadron transition density $\rho_{\rm tr}=3.4\rho_0$ for four
  values of vector coupling $G_V/G_S = 0, 0.2, 0.4, 0.6$, where $G_S$
  is the scalar coupling. The purely hadronic sequence (i.e. the
  sequence that includes nucleons and hyperons) is shown by black
  solid line.  The dashed lines and the gray solid lines show the
  branches where the 2SC and CFL quark phases are present. The filled
  circles mark the maximum masses of the sequences. The horizontal
  line shows the mass measurement of Ref.~\cite{Demorset:2010}.  {\it
    Right panel.}~ Properties of the stars as a function the free
  parameters $G_V$ and the transition density to quark matter
  $\rho_{\rm tr}$ in units of saturation density $\rho_0$. The solid
  lines show the maximum mass configurations realized for the pair of
  parameters $G_V$ and $\rho_{\rm tr}$. The dashed curves show the
  amount of CFL matter in the configurations via the ratio $\delta =
  R_{CFL}/R$, where $R_{CFL}$ is the radius of the CFL core, $R$ is
  the star radius. The parameter space to the right from $\delta = 0$
  line produces CFL stars.  The parameter space below the
  dashed-dotted curve corresponds to stars containing 2SC matter.
}
\label{fig:1}
\end{center}
\end{figure}
The mass-radius relationship for massive stars constructed on the
basis of the EoSs described above is shown in the left panel of
Fig.~\ref{fig:1} together with the mass measurement 
$M= 1.97\pm 0.04 M_{\sun}$~\cite{Demorset:2010}.  Masses above the
lower bound on the maximum mass are obtained for purely hadronic
stars; this feature is prerequisite for finding similar stars with
quark phases.  Evidently only for high values of vector coupling $G_V$
one finds stable stars that contain (at the bifurcation from the
hadronic sequence) the two-flavor superconducting (2SC) phase, which
are followed by stars that additionally contain the
color-flavor-locked (CFL) phase (for higher central densities).  Thus,
we find that the stable branch of the sequence contains stars with
quark matter in the 2SC and CFL phases.

The right panel of Fig.~\ref{fig:1} shows the changes in the masses
and composition of compact stars as the parameters of the model $G_V$
and $\rho_{\rm tr}$ are varied. First, it shows the tracks of constant
maximum mass compact stars within the parameter space. The decrease of
maximum masses with increasing vector coupling reflects the fact that
non-zero vector coupling stiffens the EoS. In other words,
to obtain a given maximum mass one can admit a small amount of soft
quark matter with vanishing vector coupling by choosing a high
transition density; the same result is obtained with a low transition
density, but large vector coupling, i.e., a stiffer quark EoS. 
For low transition densities one finds 2SC matter in stars,
which means that weaker vector couplings slightly disfavor 2SC
matter. Substantial CFL cores appear in configurations for strong
vector coupling and almost independent of the transition density
(nearly vertical dashed lines with $\delta\sim 0.1$ in
Fig. \ref{fig:1}, right panel).  Note that for a high transition
density there is a direct transition from hypernuclear to the CFL
phase. For transition densities blow $3.5\rho_0$ a 2SC layer emerges
that separates these phases. On the other hand, weak vector couplings
and low transition densities produce stars with a 2SC phase only.

\section{Thermal evolution of massive stars }
\label{sec:equilibrium}

Having developed the models of massive compact stars, we are in a
position to study their cooling, using as an input a particle
composition, that is consistent with the underlying EoS.  Such program
was carried out with the sequences of massive compact stars
constructed in Refs.~\cite{Ippolito:2007hn}.  These sequences of
stable stars permit a transition from hadronic to quark matter in
massive stars ($M> 1.85 M_{\odot}$) with the maximal mass of the
sequence $\sim 2 M_{\odot}$. Hess and Sedrakian, 
in Ref.~\cite{Alford:2004zr}, assumed quark matter
of light $u$ and $d$ quarks in beta equilibrium with
electrons.  The pairing among the $u$ and $d$ quarks occurs in two
channels: the red-green quarks are paired in a condensate with a gap of
order of the electron chemical potential; the blue quarks are
paired with a gap of order of keV, which is thus comparable to
the core temperature during the neutrino-cooling
epoch~\cite{Alford:2002rz}.  For the red-green condensate, a
parameterization of neutrino emissivity was chosen in terms of the
gaplessness parameter $\zeta = \Delta/\delta\mu$, where $\Delta$ is
the pairing gap in the red-green channel, $\delta\mu$ is the shift in
the chemical potentials of the $u$ and $d$ quarks (Jaikumar et al, in
Ref.~\cite{Alford:2004zr}).  The magnitude of the gap in the spectrum of
blue quarks was treated as a free parameter.

The stellar models were evolved in time, with the input described
above, to obtain the temperature evolution of the isothermal interior.
The interior of a star becomes isothermal for timescales $t\ge 100$
yr, which are required to dissolve temperature gradients by thermal
conduction.  Unless the initial temperature of the core is chosen too
low, the cooling tracks exit the non-isothermal phase and settle at a
temperature predicted by the balance of the dominant neutrino emission
and the specific heat of the core {\it at the exit temperature}.  The
low-density envelope maintains substantial temperature gradients
throughout the entire evolution; the temperature drops by about 2
orders of magnitude within this envelope.  The input parameters in the
evolution code are the neutrino and photon luminosities, the specific
heat of the core, and the heating processes.
\begin{figure}[t] 
\begin{center}
\includegraphics[width=12.0cm,height=8.0cm]{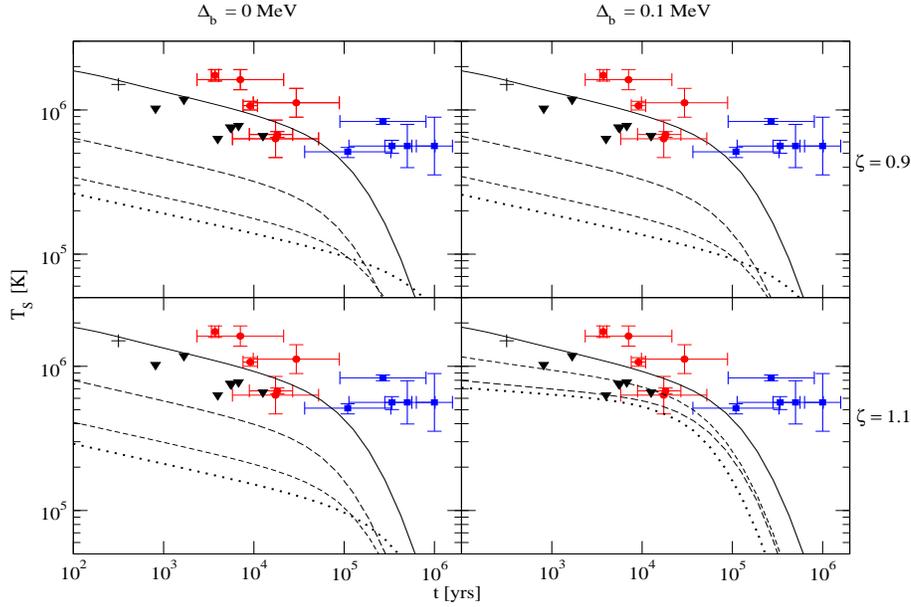}
\caption{ Time evolution of the surface temperature of four models
  with central densities   5.1
  (solid line), 10.8 (long-dashed line), 11.8 (short-dashed line),
  21.0 (dotted line) given in units of $10^{14}$ g cm$^{-3}$. 
   For the observational data see Hess and Sedrakian in Ref.~\cite{Alford:2004zr}.
  The upper two panels correspond to cooling when the red-green
  condensate has $\zeta = 0.9$, i.e., is not fully gapped; the lower
  panels correspond to $\zeta = 1.1$, i.e., the red-green condensate is
  fully gapped. The left two panels correspond to evolution with
  negligible blue-quark pairing ($\Delta_b =0$); the right two panels
  show the evolution for large blue pairing $\Delta_b=0.1$~MeV.  }
\label{fig:3}
\end{center}
\end{figure}
The resulting evolution tracks are shown in Fig.~\ref{fig:3}, where we
display the dependence of the (redshifted) surface temperature on
time. Each panel of Fig.~\ref{fig:3} contains cooling tracks for the
same set of four models with increasing central density. The cooling
tracks for the purely hadronic model (solid lines) are the same in all
four panels. The panels differ in the values of micro-physics
parameters (for further explanations see the caption of 
Fig.~\ref{fig:3}).  It can be seen that (i) the neutrino-cooling is
slow for hadronic stars and becomes increasingly fast with an increase
of the size of the quark core, in those scenarios where there are
unpaired quarks or gapless excitations in the superconducting quark
phase. The temperature scatter of the cooling curves in the neutrino
cooling era is significant and can explain the observed variations in
the surface temperature data of same age neutron stars. (ii)~If quarks
of all colors have gapped Fermi surfaces, the neutrino cooling shuts
off early, below the pairing temperature of blue quarks; in this case,
the temperature spread of the cooling curves is not as significant as
in the fast cooling scenarios. (iii) As the stars evolve into the
photon cooling stage the temperature distribution is inverted, i.e.,
those stars that were cooler in the neutrino-cooling era are hotter
during the photon cooling stage.

Our preliminary studies show that the rapid cooling of Cas A can be
understood as an evidence of a phase transition from perfect 2SC phase
to a crystalline color superconducting state. We assume the
Fulde-Ferrell type superconductivity as discussed in
Refs.~\cite{Sedrakian:2009kb}; (for more complex order parameters
see Ref.~\cite{Mannarelli:2007bs} and references therein).  The phase
transition is characterized by its temperature $T^*$ and the timescale
$\tau^*$ needed to complete such a phase transition. While {\it
  equilibrium }model calculations provide a value for $T^*\sim 10$~MeV
in a particular (NJL Lagrangian based) model, the physically
realizable value of $T^*$ may depend on dynamical factors and
nucleation history of the quark phase.  Because of the density
dependence of the critical temperature $T^*$ in the stellar interiors,
this phase transition will not occur globally in the quark core, but
gradually within some shells. Therefore, the values of $T^*$ and
$\tau^*$ will be determined by the local conditions in the shells, the
transition being a gradual propagation of phase-separation fronts
rather than a coherent instantaneous transition.  In our study we
treat the transition temperature $T^*$ as the key free parameter to
fit the cooling model to the Cas A data. These fits are not sensitive
to the precise value of time $\tau^*$ characterizing the duration of
the transition.

\begin{figure}[tb]
\begin{center}
\vskip 0.2cm
\includegraphics[width=13.cm,height=7cm]{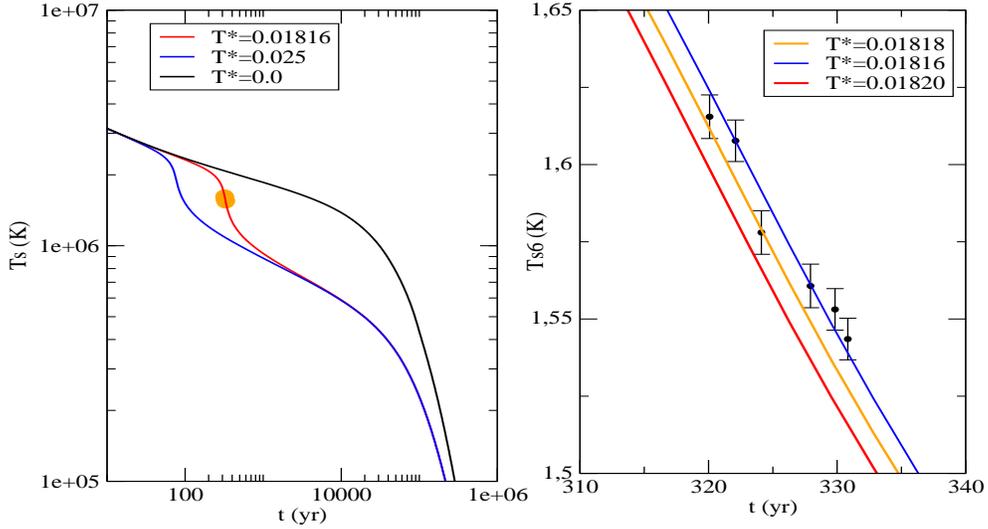}
\caption{{\it Left panel.} Dependence of the surface temperature on
 time for the values of the transition temperature $T^*$ (in MeV) indicated in
 the plot and for fixed $\tau^*$. The underlying model is the 1.91
 $M_{\odot}$ compact star model from Ref.~\cite{Ippolito:2007hn}.
 {\it Right panel.}  Dependence of the surface temperature (in units
 of $10^6$ K) on time for the same model.  The points with error bars
 show the Cas A data, the solid lines are fits to these data by
 variation of $T^*$. }
\label{fig2}
\end{center}
\end{figure}

Fig.~\ref{fig2}, left panel, shows the cooling evolution of the $M =
1.91 M_{\odot}$ model of a compact star for several values of $T^*$
and fixed $\tau^*$. It is seen that for a specific value of $T^*$ the
cooling curves pass through the location of Cas A in the log $T$-log
$t$ plain.  The right panel of Fig.~\ref{fig2} shows the
observationally covered evolution of Cas A for the same value of
$\tau^*$ and several (fine-tuned) values of $T^*$.  It is seen that
the data can be accurately fitted by fine-tuning a single parameter
$T^*$. In each case the critical temperature of pairing for the blue
quarks is assumed to satisfy the condition $T_{cb} \gg T^*$, which
implies that blue quarks have no effect on the dynamics of the
transition process. The numerical value $T_{cb} = 0.1$ MeV was used in
the simulations.

\section{Color-magnetic flux tubes }

We turn now to the response of dense quark matter to a magnetic
field. Our discussion will be focused on the formation and dynamics of
color-magnetic flux tubes in the two-flavor color superconducting (2SC) phase.
Type-II superconductivity leads to formation of flux tubes if the cost
of inserting a flux tube, which has always a positive self-energy
(flux-tension) in a superconductor is compensated by the interaction
energy of the flux tube with external magnetic field intensity. The
critical intensity (lower critical field) $H_{c1}$ for the formation
of Abrikosov flux tubes containing color-magnetic flux is large, of
order $10^{17}$ G.  However, when the quark matter cools into the 2SC
phase, the process of domain formation and amalgamation is likely to
leave some of the flux trapped in the form of flux
tubes~\cite{Alford:2010qf}. These flux-tubes may be topologically
unstable, however their formation and decay timescales in 
an externally imposed field are unknown. 

The 2SC phase contains three species of gapless fermions: two quarks
(``blue up'' and ``blue down'') and the electron.  These are expected
to dominate its transport properties.  Light ungapped fermions will be
scattered from color-magnetic flux tubes due to the Aharonov-Bohm
(hereafter AB) effect~\cite{Alford:1988sj}.  The scattering of
electrons and blue quarks of the flux tubes in the 2SC cores of
compact stars contributes to the transport coefficients of matter and
acts as a dissipative force on the flux tubes. The relaxation time for
incident light fermion on a flux tube is given by~\cite{Alford:2010qf}
$ \tau^{-1}_{f} = (n_v/p_{F}) \sin^2(\pi\tilde\beta), $ where $n_v$ is
the flux density, $p_{F}$ is the fermion momentum and $\beta$ is the
dimensionless AB parameter computed in Ref.~\cite{Alford:2010qf}. It
is easy to understand this result. It is of the standard form for
classical gases $\tau^{-1}=c n \sigma$, where $c=1$ is the speed of
the particles, $n=n_v$ is the density of scattering centers, and
$\sigma\propto \sin^2(\pi\tilde\beta)/p_F$ is the cross section for AB
scattering.  One of the blue quarks, the blue down quark, has no AB
interaction with the flux tubes ($\tilde\beta=0$). The other two light
fermions, the electron and blue up quark, have identical AB factors
although their Fermi momenta are different.

Because the ambient magnetic field in a neutron star is below the
lower critical field required to force color-magnetic flux tubes into
2SC quark matter, the trapped flux tubes will feel a boundary force
pulling them outwards.  This force will be balanced by the drag force
(``mutual friction'') on the moving flux tube due to its AB
interaction with the thermal population of gapless quarks and
electrons, and also by the Magnus-Lorentz force.  An estimate of the
timescale for the expulsion of color-magnetic flux tubes from a 2SC
core on the basis of balance of forces acting on the vortex shows that
it is of order $10^{10}$ years~\cite{Alford:2010qf}.  Therefore, it is
safe to assume that the magnetic field will be trapped in the 2SC core
over evolutionary timescales in the absence of other external forces.

\section{Final remarks}

Much remains to be done. First, our understanding of the EoS of dense
quark matter and its phases needs further exploration.  The
strangeness degrees of freedom including hypernuclear matter and
three-flavor quark matter should be further explored building, e.g.,
upon the work of Ref.~\cite{Bonanno:2011ch}. The equilibrium and
stability of massive compact objects, constructed from these EoSs,
should be studied including rapid rotations and oscillations.  Second,
we need a better understanding of the weak interaction rates in quark
and (hyper)nuclear matter, which are required input in cooling
simulations of compact stars. Third, the transport coefficients of
dense color superconducting quark matter, such as the thermal
conductivity, are input for modelling an array of phenomena, which
include thermal evolution, magnetic evolution, and oscillation modes.

I would like to thank M.~Alford, L.~Bonanno, D.~Hess, X.-G. Huang,
D.~Rischke for the collaboration on the topics discussed in this
contribution.


\begin{thebibliography}{99}
\bibitem{Kramer:2012zz} 
  M.~Kramer,
  arXiv:1211.2457 [astro-ph.HE].

\bibitem{Demorset:2010} 
P. B. Demorest, et al.
Nature {\bf 467}, 1081 (2010).

\bibitem{MassiveStars}
  K.~Masuda, T.~Hatsuda and T.~Takatsuka,
  arXiv:1212.6803 [nucl-th];
  P.~-C.~Chu and L.~-W.~Chen,
  arXiv:1212.1388 [astro-ph.SR];
  C.~H.~Lenzi and G.~Lugones,
  Astrophys.\ J.\  {\bf 759}, 57 (2012);
  J.~L.~Zdunik and P.~Haensel,
  arXiv:1211.1231 [astro-ph.SR].
  V.~Dexheimer, J.~Steinheimer, R.~Negreiros and S.~Schramm,
  arXiv:1206.3086 [astro-ph.HE];
  R.~Lastowiecki, D.~Blaschke, H.~Grigorian and S.~Typel,
  Acta Phys.\ Polon.\ Supp.\  {\bf 5}, 535 (2012);
  A.~Sedrakian,
  Acta Phys.\ Polon.\ B {\bf 5}, 867 (2012);
  N.~K.~Johnson-McDaniel and B.~J.~Owen,
  Phys.\ Rev.\ D {\bf 86}, 063006 (2012);
  J.~M.~Lattimer and M.~Prakash,
  arXiv:1012.3208 [astro-ph.SR];
  A.~Sedrakian,
  AIP Conf.\ Proc.\  {\bf 1317}, 372 (2011)
See also 
M.~Alford, M.~Braby, M.~W.~Paris and S.~Reddy,
Astrophys.\ J.\  {\bf 629}, 969 (2005);
A.~Kurkela, P.~Romatschke, A.~Vuorinen and B.~Wu,
arXiv:1006.4062 [astro-ph.HE].




\bibitem{Bonanno:2011ch} 
 L.~Bonanno and A.~Sedrakian,
 Astron.\ Astrophys.\  {\bf 539}, A16 (2012).


\bibitem{Radii}
  S.~Guillot, R.~E.~Rutledge and E.~F.~Brown,
  Astrophys.\ J.\  {\bf 732}, 88 (2011);
  S.~Bogdanov,
  arXiv:1211.6113 [astro-ph.HE];
  D.~K.~Galloway and N.~Lampe,
  Astrophys.\ J.\  {\bf 747}, 75 (2012);
  T.~Guver, P.~Wroblewski, L.~Camarota and F.~Ozel,
  Astrophys.\ J.\  {\bf 719}, 1807 (2010).



\bibitem{Alford:2004zr}
M.~Alford, P.~Jotwani, C.~Kouvaris, J.~Kundu and K.~Rajagopal,
Phys.\ Rev.\  D {\bf 71}, 114011 (2005);
  H.~Grigorian, D.~Blaschke and D.~Voskresensky,
  Phys.\ Rev.\  C {\bf 71}, 045801 (2005);
 P.~Jaikumar, C.~D.~Roberts and A.~Sedrakian,
 Phys.\ Rev.\ C {\bf 73}, 042801 (2006).
  R.~Anglani, G.~Nardulli, M.~Ruggieri and M.~Mannarelli,
  Phys.\ Rev.\  D {\bf 74}, 074005 (2006);
  D.~Hess and A.~Sedrakian,
  Phys.\ Rev.\ D {\bf 84}, 063015 (2011);
  T.~Noda, M.~-a.~Hashimoto, Y.~Matsuo, N.~Yasutake, T.~Maruyama, T.~Tatsumi and M.~Fujimoto,
  arXiv:1109.1080 [astro-ph.SR];
  N.~Yasutake, T.~Noda, H.~Sotani, T.~Maruyama and T.~Tatsumi,
  arXiv:1208.0427 [astro-ph.HE].



\bibitem{Shternin:2010qi}
P.~S.~Shternin, et al.
Mon.\ Not.\ Roy.\ Astron.\ Soc.\  {\bf 412}, L108 (2011).

\bibitem{QV}
  D.~Blaschke, D.~M.~Sedrakian and K.~M.~Shahabasian,
  Astron.\ Astrophys.\  {\bf 350}, L47 (1999).
  D.~M.~Sedrakian, D.~Blaschke, K.~M.~Shahabasyan and D.~N.~Voskresensky,
  Astrofiz.\  {\bf 44}, 443 (2001)
  K.~Iida and G.~Baym,
  Phys.\ Rev.\ D {\bf 65}, 014022 (2002);
  K.~Iida and G.~Baym,
  Phys.\ Rev.\ D {\bf 66}, 014015 (2002);
  K.~Iida and G.~Baym,
  Phys.\ Rev.\ D {\bf 63}, 074018 (2001)
  [Erratum-ibid.\ D {\bf 66}, 059903 (2002)];
K.~Iida, Phys.    Rev. {\bf D71}, 054011 (2005);
A.~P. Balachandran, S.~Digal, and T.~Matsuura, 
Phys. Rev. {\bf D73},  074009 (2006);  D. M. Sedrakian et al., 
Astrophys. {\bf 51},  544 (2008).


\bibitem{Alford:2010qf}
  M.~G.~Alford and A.~Sedrakian,
  J.\ Phys.\ G {\bf 37}, 075202 (2010).



\bibitem{CSCGeneral}
 M.~G.~Alford, et al.
 Rev.\ Mod.\ Phys.\  {\bf 80}, 1455 (2008);
  Q.~Wang,
  Prog.\ Phys.\  {\bf 30}, 173 (2010);



\bibitem{LBL-30645} 
  N.~K.~Glendenning and S.~A.~Moszkowski,
  Phys.\ Rev.\ Lett.\ \ {\bf 67}, 2414  (1991).

\bibitem{Ippolito:2007hn}
  N.~Ippolito, et al.
  Phys.\ Rev.\  D {\bf 77}, 023004 (2008); 
  B.~Knippel and A.~Sedrakian,
  Phys.\ Rev.\  D {\bf 79}, 083007 (2009).

\bibitem{Alford:2002rz}
 M.~G.~Alford, J.~A.~Bowers, J.~M.~Cheyne and G.~A.~Cowan,
 Phys.\ Rev.\  D {\bf 67}, 054018 (2003);
M.~Buballa, J.~Hosek and M.~Oertel,
Phys.\ Rev.\ Lett.\  {\bf 90}, 182002 (2003);
 A.~Schmitt, Q.~Wang, and D.~H. Rischke, Phys. Rev. Lett. {\bf 91},
  242301 (2003); 
A.~Schmitt,
Phys.\ Rev.\  D {\bf 71}, 054016 (2005).


\bibitem{Sedrakian:2009kb} 
 A.~Sedrakian and D.~H.~Rischke,
 Phys.\ Rev.\ D {\bf 80}, 074022 (2009);
  X.~-G.~Huang and A.~Sedrakian,
  Phys.\ Rev.\ D {\bf 82}, 045029 (2010).


\bibitem{Mannarelli:2007bs} 
 M.~Mannarelli, K.~Rajagopal and R.~Sharma,
 Phys.\ Rev.\ D {\bf 76}, 074026 (2007).

\bibitem{Alford:1988sj}
M.~G. Alford and F.~Wilczek, 
Phys. Rev. Lett. {\bf 62}, 1071 (1989).

\end{thebibliography}
\end{document}